# Interest Rate Manipulation Detection using Time Series Clustering Approach

Murphy Choy
Koo Ping Shung
Enoch Chng

Abstract

The Interbank Offered Rate is a vital benchmark interest rate in the financial markets of every country to which financial contracts are tied. In the light of the recent LIBOR manipulation incident, this paper seeks to address the fear that Interbank Offered Rate are entirely controlled by the bank. The paper will focus on the comparison between LIBOR and SIBOR especially with regards to the behavior of the interest rate with time. Because of the nature of IBORs, banks will naturally be submitting similar rates which should not differ excessively from the market as well as the other banks. We will compare the LIBOR and SIBOR from 2005 to 2011 with respect to the 1 month rates on an annual basis. We will present the result that the SIBOR is not manipulated like LIBOR.

## Introduction

The LIBOR scandal first erupted in 2008 at the height of the 2007-2009 Global Financial crisis. Wall Street Journal first released a study indicating that the banks were deliberately understating the rates in an attempt to improve their financial positions. Both the BIS and BBA responded with statements indicating the reliability of the rates quoted citing the difficulties in the financial markets for the discrepancies in the rates published. This position is further supported by the IMF's regular reports. However, an independent group of researchers(Snider and Youle, 2010) did find results which corroborated with the Wall Street Journal article. However, the researchers believe that the banks were attempting to profit from the movements in the rates rather than strengthening the banks' positions.

With the new results, financial and fraud investigation commenced on several international banks including Barclays which is the first bank to admit to LIBOR fixing. This investigation rapidly spread to other financial markets such as Canada and Singapore. Many issues and concerns were raised by the financial regulators about the importance of separation of powers and whether new cross-boundary regulations should be implemented to prevent a repeat of the incident. Others seek a complete overhaul fo the IBOR calculation framework in hope that it will prevent manipulations. In the following section, we will be demonstrating how the cluster analysis of time series can be used to detect such irregularities and how the IBORs are currently affected by the manipulations.

## Time Series Clustering

While time series data incorporate time as a dimension, performing clustering on time series is similar to the clustering on static data. The technique employed to form clusters depends on the type of data available as well as the purpose and application of the analysis. Given the nature of time series, most time series data are continuous values and univariate in nature.

While there are several algorithms developed solely for the purpose of clustering time series data of different nature, they are all similar in their approach of modifying existing non time series data algorithms to manage time series. This approach assumes raw time series data otherwise also known as raw-data-based approach and normally involves data pre-processing to covert the time series data to a normal static form. Once the data has been pre-processed, the classical static data clustering algorithms can then be applied.

Every clustering algorithms relies on a suitable measure to compute either distance or similarity between two time series. Certain particular measure might be more appropriate than another depending on the type of time series in question. Most clustering algorithms are iterative in nature and rely on a suitable criterion to determine whether clustering obtained is in good condition to stop the iterative process. There were extensive discussions about the problems of using the Euclidean distance in the comparison of time series (Keogh, Lin and Truppel, 2003). However, most of the criticisms are targeted at the pattern recognition of time series behaviors. Subsequent research indicated that in certain cases where pure pattern recognition is not the main objective, direct application of the Euclidean distance measure as the main clustering input does not affect the effectiveness of the technique (Chen, 2005).

The hierarchical clustering method works by grouping data objects into a tree of clusters. There are two types of hierarchical clustering methods. The two major clustering process are agglomerative or divisive depending upon whether the bottom-up or top-down strategy is followed. The agglomerative approach is more commonly used than the divisive method. The algorithm starts by having each object in its own cluster and start merging the atomic clusters into larger and larger clusters until all the objects are in a single cluster. The single linkage approach measures the similarity between two clusters using the similarity of the closest pair of data points between the clusters. The closest two clusters are merged and the process repeats merging until all the clusters forms one cluster. The Ward's minimum variance approach differs by merging the two clusters that minimally increase the value of the sum-of-squares variance. At every step, all possible mergers of two clusters are tried and the one with the smallest increase is selected. The agglomerative hierarchical clustering method often suffers from adjustment problem. Hierarchical clustering is not restricted to cluster time series with equal length and can be extended to series of unequal length with the appropriate distance measure.

**IBOR rate mechanism**

The IBOR rate mechanism is extremely similar across the various countries and jurisidictions. The most common cited IBOR is the LIBOR which was standardized in 1984 by the British Bankers' Association as the main reference rate for numerous securities which includes syndicated loans, futures contracts and forward rate agreements. The LIBOR is also used to as the reference for unsecured instruments between banks in London and globally. Globally, all the IBORs behave similarly to LIBOR in that they are quoted daily for several major currencies.

The various IBORs rely on a panel of selected banks to provide daily rate qoutes for the calculation of the IBOR. The banks are selected based a variety of criteria such as size of operation, reputation as well as capabilities and knowledge of the currency concerned. Typically, the biggest banks operating in the particular currency will be consulted for the rates.

Currently, Libor is defined as the rate at which an individual Contributor Panel bank could borrow funds, were it to do so by asking for and then accepting inter-bank offers in reasonable market size, just prior to 11.00 London time.(BBA, 2012)

This definition is further broken down into the following sections (BBA, 2012):
- •The submitted rate must be formed from that bank's estimated cost of funds in the interbank market.
- •Contributions must represent rates formed in London only.

- Contributions addressed only the currency concerned and does not seek to address the cost of producing one currency by borrowing in another currency and accessing the required currency via the foreign exchange markets.
- The rates must be submitted by members of staff at a bank with charged with the management of a bank's cash.
- The "funds" is defined as the unsecured interbank cash or cash raised via the issuance of interbank Certificates of Deposit.

Every Panel Bank is required to directly input its data no later than a given time for each day that the capital market is open. After the given time, an appointed agent will then process the calculation. Usually, for each maturity, the agent will eliminate the highest and lowest X% of all the quotes collected to eliminate the outliers. The remaining rates will be averaged and rounded to three decimal places. It is precisely this approach that resulted in the 2012 LIBOR Scandal.

**LIBOR Scandal**

The Libor scandal is a series of events that consist of fraudulent actions by the bank with regards to their behaviors with regards to the LIBOR. The incidents arose when the banks' action of falsely inflating or deflating their rates so as to profit from trades and implant false impression of creditworthiness.

As mentioned in the earlier section, the banks are supposed to submit the actual interest rates they are paying (or based on their expectation) for borrowing from other banks. The Libor behaves like an overall assessment of the health of the financial system and acts as the litmus test for financial health of the banks. If the banks being polled feel confident about the state of affair, they will report a low number and vice versa in a situation of low confidence.

As Libor is used in U.S. derivatives markets as a benchmark, any attempt to manipulate Libor is considered to be an attempt to manipulate U.S. derivatives markets and violates the American laws. Since many financial products rely on Libor as the reference rate, any manipulation of the submissions for the calculation of the rates can have strong and significant negative impacts on consumers and financial markets worldwide.

There were strong debates about how the submissions could have affected the LIBOR. Lively discussions erupted online and two major camps formed. The first group of thinkers believed that the nature of the LIBOR calculation makes it impossible for any one single bank to manipulate it and any manipulations must be the product of several banks colluding with one another(Persaud, 2012). The second group of thinkers believes that the LIBOR has been manipulated and that any bank can manipulate the rates (Smith, 2012).

From the discussions, there are several key items that will mentioned and discussed. In the following section, we will be addressing these items. The items that need to be answered are as follow:

1. Can LIBOR be manipulated by 1 bank?
2. Were the banks involved in LIBOR colluding? If so, who?
3. Was SIBOR similar?

**Experiment**

To detect irregularities in the IBOR rates, we will first extract data from suitable sources. To achieve this, the authors have extracted the LIBOR and SIBOR submission data from Bloomberg data services. To cover the period when there were data manipulation, data for LIBOR and SIBOR is collected from 2005 to 2012 for 1 Month maturity. This is to ensure consistencies in comparison.

To answer whether LIBOR can be manipulated by a single bank, we have to first understand the mechanism of the LIBOR calculation which is based on the concept of trimmed mean. The trimmed mean is a type of statistical measure of central tendency which is similar to the mean and median. It is calculated by removing both ends of the extreme values. Let us look at the following example.

Assuming that we have 10 banks, and under normal circumstances, we have the rate submitted below.

| Bank | Rates |
|------|--------|
| 1 | 3.0026 |
| 2 | 3.0106 |
| 3 | 3.0235 |
| 4 | 3.0312 |
| 5 | 3.0358 |
| 6 | 3.0434 |
| 7 | 3.0562 |
| 8 | 3.0601 |
| 9 | 3.0658 |
| 10 | 3.0961 |
| LIBOR | 3.04168 |

Table 1: Theoretical LIBOR Scenario(No Manipulations)

Let us assumes that Bank 9 wishes to lower LIBOR and submits a very low rate. Below is the scenario,

| Bank | Rates |
|------|--------|
| 1 | 3.0026 |
| 2 | 3.0106 |
| 3 | 3.0235 |
| 4 | 3.0312 |
| 5 | 3.0358 |
| 6 | 3.0434 |
| 7 | 3.0562 |
| 8 | 3.0601 |
| 9 | 3.0000 |
| 10 | 3.0961 |
| LIBOR | 3.0334 |

Table 2 Theoretical LIBOR Scenario(With manipulations)

From the example, we can see that the LIBOR has been successfully been lowered. Thus the answer to question 1 and 2 is clear, any single bank can through its submission of rates manipulate the rates without having to even collude with any other banks. Thus it is clear that the existing approach can be vulnerable to manipulation.

To prevent any manipulations of the IBORs, the only possible measure is to attempt to detect anomalous behavior in the rate submissions. This is considered simple to achieve by conducting time series clustering on the data. This is because any banks attempting to manipulate the rates will display behaviors which are distinctively different from the rest of banks. Early splits in the dendrogram of cluster analysis will reveal unique clusters of banks which behaves very differently from others. This can be seen from the results in the next section.

**Results**

We first begin our analysis on the 2005 and 2008 data where there were previous studies that Barclays has actually attempted to manipulate the rates. Below is the 2005 cluster analysis results.

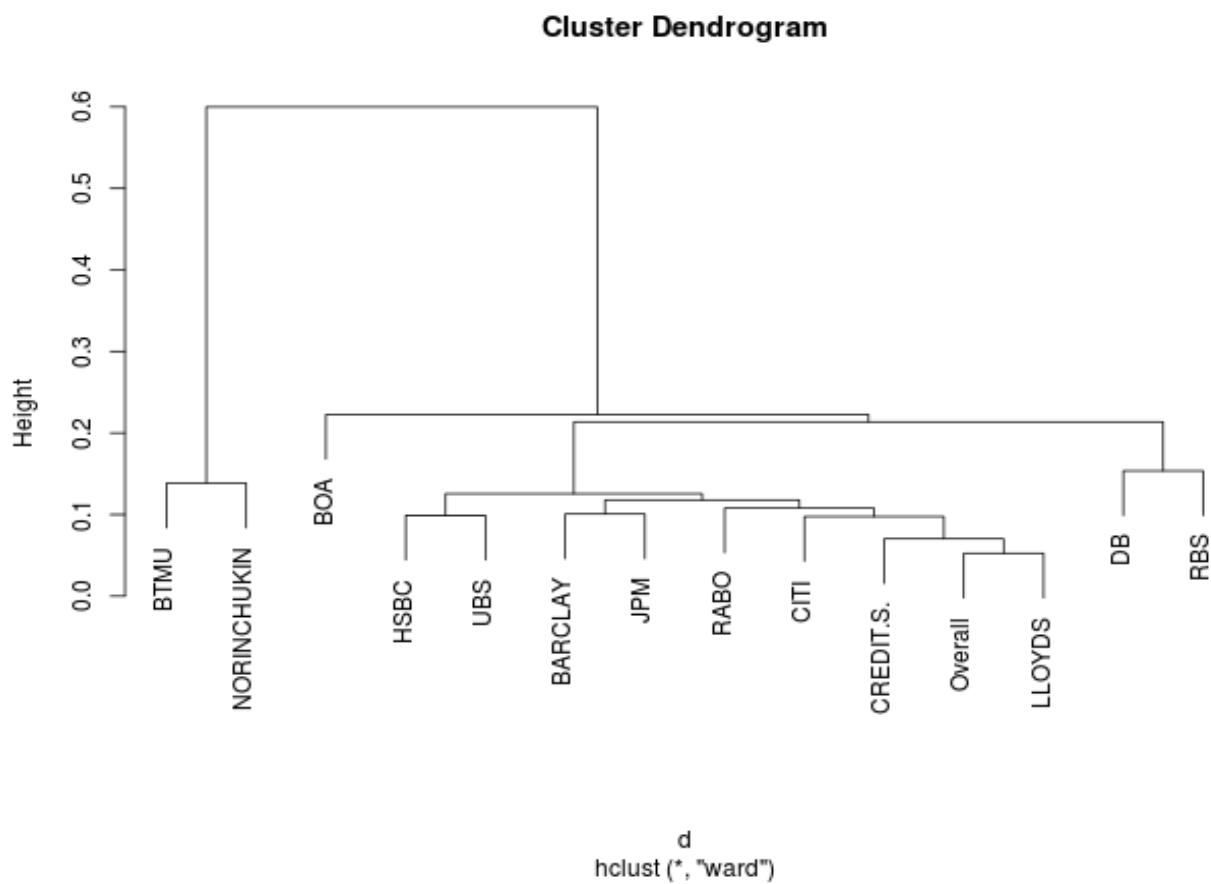

Chart 1: LIBOR 2005 Cluster Analysis

From the chart, we can observe that there are several distinct groups. However, Barclays does not display any significant deviation from the rest of the banks for 2005 for 1 month maturity. The most distinctive groups are the BTMU/NORIN, BOA and DB/RBS. From the distance measured, it is clear that most of the rates are quite similar to one another. However, the picture could not have been more distinct in the year of 2008 as shown below.

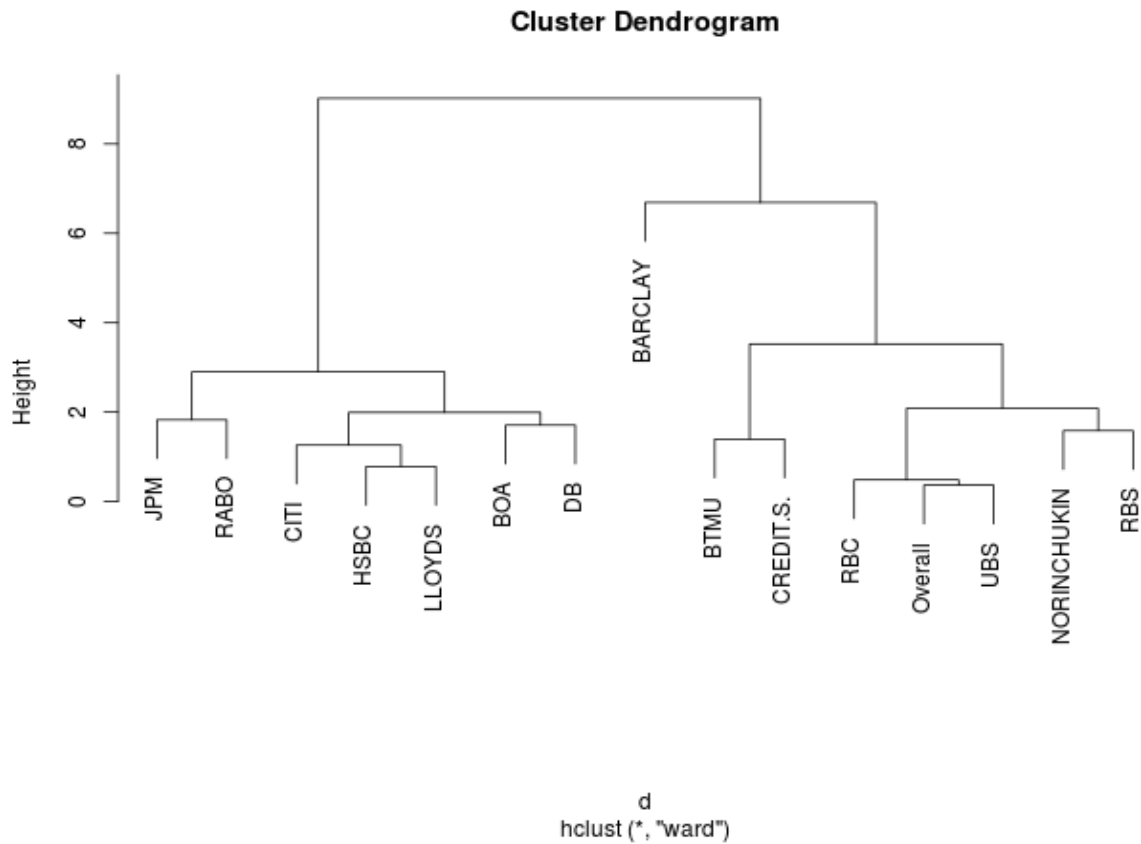

Chart 2: LIBOR 2008 Cluster Analysis

From the analysis, we can observe that Barclays display significant deviation from the rest of the banks for 2008 for 1 Month maturity. While there are two major groups of banks, they are distinctively different from one another and can be attributed to the financial distress from the crisis. Banks such as Citibank and JP Morgan were under less severe market conditions as compared to Credit Suisse as they were assisted by the TARP. However, we can see that Barclays is definitely different from the rest of the banks as shown below in table 3. Thus for question 2, there was no obvious collusion betweent the banks.

| Banks | Rates |
|---|---|
| JPM | 2.600 |
| RABO | 2.607 |
| DB | 2.616 |
| CITI | 2.643 |
| BOA | 2.648 |
| LLOYDS | 2.653 |
| HSBC | 2.656 |
| UBS | 2.673 |
| RBC | 2.673 |
| Overall | 2.674 |
| RBS | 2.686 |
| CREDIT S. | 2.727 |
| NORINCHUKIN | 2.727 |
| BTMU | 2.738 |
| BARCLAY | 2.783 |

Table 3: LIBOR Rates 2008 (Average Daily)

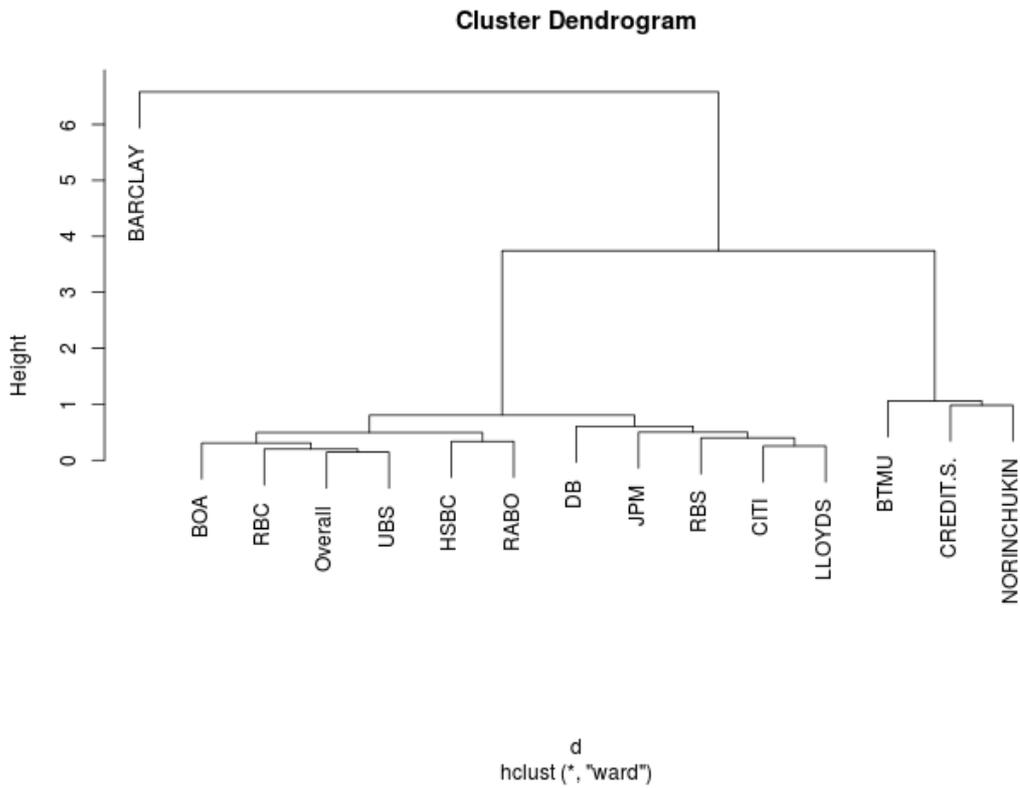

Chart 3: LIBOR 2008 Quarter 3 Cluster Analysis

The cluster analysis result is also verified by examining the data from 2008 quarter 3 where Barclays has behaved differently from the rest of the banks. We can see that almost all the banks are similar compared to Barclays indicating that there were certain actions which were anomalous. Let us review the SIBOR for 2007 and 2008 financial crisis period.

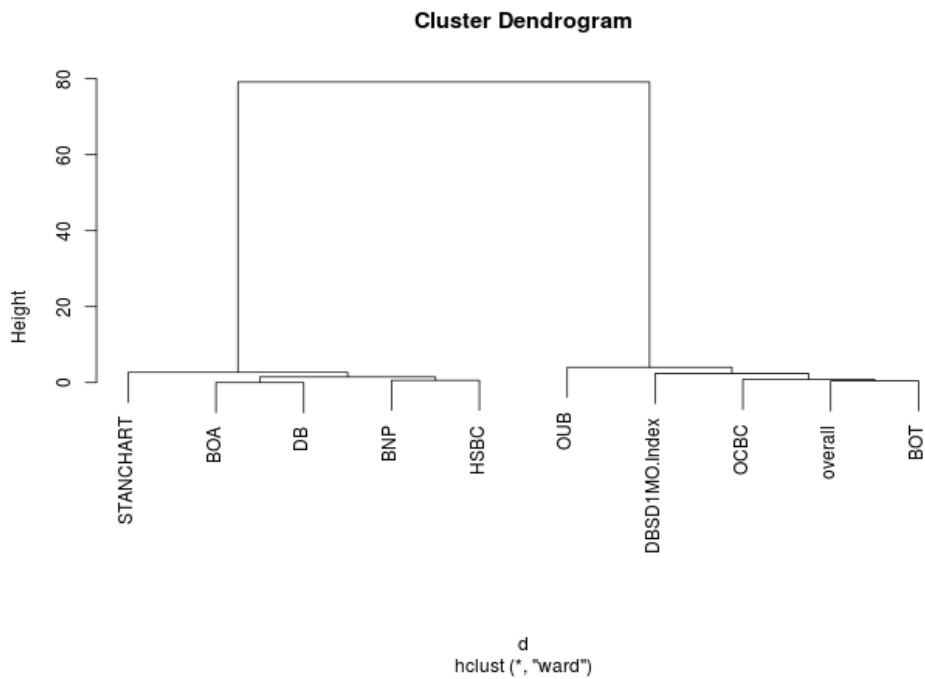

Chart 4: SIBOR 2007 Cluster Analysis

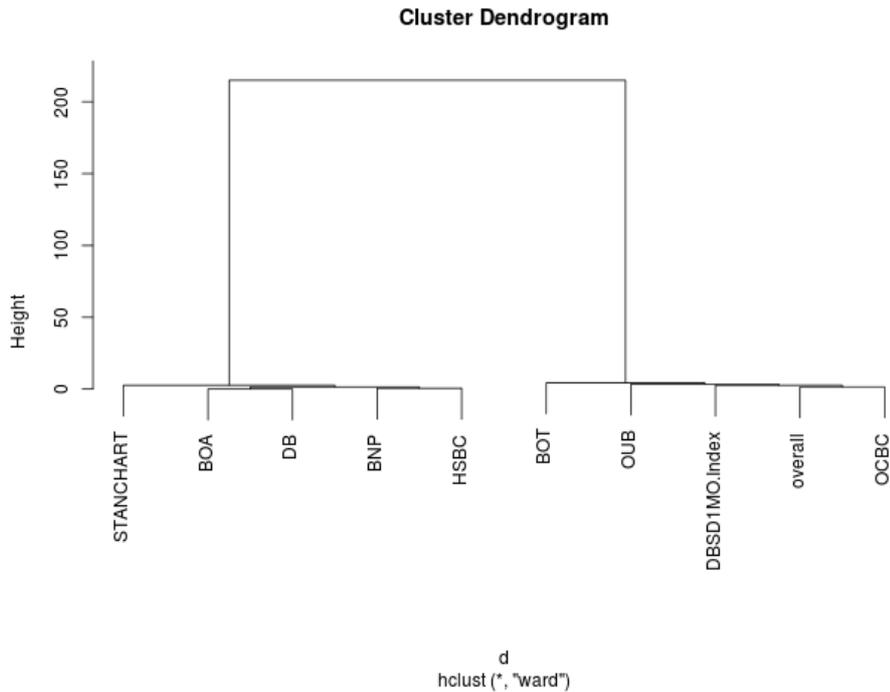

Chart 5: SIBOR 2008 Cluster Analysis

The SIBOR charts (4,5) display information that indicates that there are no distinct individual which is very different from the rest of the banks. There are two distinct groups. The first group comprises of the major local banks while the second group consists of overseas or international banks.

| Banks | Rates |
|---|---|
| DBSD1MO Index | 2.627 |
| OUB | 2.648 |
| overall | 2.658 |
| OCBC | 2.662 |
| BOT | 2.668 |
| STANCHART | 3.375 |
| BNP | 3.438 |
| HSBC | 3.469 |
| BOA | 3.500 |
| DB | 3.500 |

Table 4: SIBOR Rates 2007 (Average Daily)

| Banks | Rates |
|---|---|
| DBSD1MO Index | 1.070 |
| OCBC | 1.130 |
| overall | 1.152 |
| OUB | 1.156 |
| BOT | 1.211 |
| STANCHART | 3.375 |
| BNP | 3.438 |
| HSBC | 3.469 |
| BOA | 3.500 |
| DB | 3.500 |

Table 5: SIBOR Rates 2008 (Average Daily)

The major reason for the differences is the financial distress experienced by the international banks in their home country lending markets. Another reason is the lack of exposure of the local banks to the toxic portfolio components of subprime lending. There are not evidence that they were attempting to manipulate the rates by colluding with one another. Thus, for question 3, we can safely say that SIBOR is not manipulated during the financial crisis.

**Conclusion**

From the analysis, the cluster analysis technique was able to identify Barclays as the bank behaving anomalously from the rest of the bank. The dendrogram identifies the most obvious bank which is not behaving according to market behavior. In the LIBOR case for 2005 and 2007, there were no obvious collusions with the exception of Barclays. SIBOR was not affected by rate manipulation between 2007 and 2008. However, the technique has some weaknesses as well.

The cluster analysis while being able to detect rogue behaviors, is unable to identify collusions that are extremely well planned and affects the rates mildly. As demonstrated in the sections above, any bank can influence the rates. Suppose several banks wishes to affect the rate by 0.01%, that can be done easily by quoting the same lowest rate together and thus forming a group. In the dendrogram, they will be found as a group. The group will be so big that they will be similar to the SIBOR case. Although further investigation into the manner of grouping might reveal anomalies, it cannot be directly inferred from the diagram. In the SIBOR case, the grouping is obvious due to the origin of the banks which renders addition investigation uncalled for. Further research will be needed to enhance the analysis so that the display of collusion can be incorporated into the analysis. Further investigation will also be needed to understand how many banks colluding together will nullify the technique as well.